\documentclass[%
 aip,
 jmp,%
 amsmath,amssymb,
reprint,%
]{revtex4-1}

\usepackage{hyperref}
\usepackage{cleveref}
\usepackage{color}
\usepackage{graphicx}
\usepackage{dcolumn}
\usepackage{bm}

\usepackage{epstopdf}

\begin{document}


\title{Characterization of the continuous transition from atomic to molecular shape in the three-body Coulomb system}

\author{Laura D. Salas}

\author{B\'arbara Zamora-Yusti}
\altaffiliation{Present address: Department of Inorganic and Analytical Chemistry, and MTA-BME
Computer Driven Chemistry Research Group,
Budapest University of Technology and Economics,
Szent Gell\'ert t\'er 4, 1111 Budapest, Hungary}

\author{Julio C. Arce}
\email{julio.arce@correounivalle.edu.co}
\affiliation{Departamento de Qu\'imica, Universidad del Valle, A.A. 25360, Cali, Colombia
}%

\date{\today}

\begin{abstract}

\noindent We present an alternative, univocal characterization of the continuous transition from atomic to molecular shape in the Coulomb system constituted by two identical particles and a third particle with the opposite charge, as the mass ratio of the particles varies. Applying a marginal-conditional exact factorization to a variationally optimized wavefunction, we construct a nonadiabatic potential energy surface for the relative motion between the single particle and each of the identical particles in the ground state. The transition is revealed through the evolution with the mass ratio of the topography of such surface and of the shapes of the associated marginal and conditional distributions. Our approach unifies and extends to the nonadiabatic regime the Born-Oppenheimer and charge-distribution pictures of molecular shape.

\end{abstract}


\maketitle

\section{\label{sec:introduction} Introduction}


At present, thanks mainly to Woolley's early discussion,
it is widely recognized that the justification of 
the so-called molecular structure hypothesis, i.e., the paradigm of modeling an isolated molecule as a semi-rigid three-dimensional array of atoms held together by chemical bonds, inherited from the classical structural theory of chemistry, is problematic for a fully ab-initio quantum-mechanical theory of molecules \cite{Woolley1976,Woolley1978,Wilson1979,Claverie1980,Garcia1981,Primas1983,Weininger1984,Woolley1985,Woolley1986,Woolley1988a,Woolley1988b,Sutcliffe1992,Sutcliffe2012,Sutcliffe2013,Mills2018,Martinez2019,Sutcliffe2020}.
This problem stems from the facts that the molecular Hamiltonian is entirely and uniquely specified just by the masses, charges and spins of all the constituent nuclei and electrons,
without any a priori reference to atoms, bonding or geometry, and that the molecular wavefunction must exhibit the appropriate permutation symmetry for all subsets of indistinguishable particles \cite{Woolley1976,Woolley1988a,Sutcliffe1992,Sutcliffe2012,Sutcliffe2013,Sutcliffe2020,Bunker1998}.
Thus, prima facie to recover such classical-like description of molecular structure auxiliary considerations must be added to the fully quantal treatment \cite{Woolley1976,Woolley1978,Claverie1980,Woolley1986,Woolley1988a,Sutcliffe2012,Sutcliffe2013,Sutcliffe2020}.

In quantum chemistry, the notion of a classical-like atomic arrangement, commonly called the `molecular shape' \cite{Woolley1978,Woolley1988b}, is recovered from the topography of the Born-Oppenheimer (BO) potential energy surface (PES) \cite{Sutcliffe2012,Sutcliffe2013}, whose isolated minima \emph{define} the geometries of the nuclear frameworks of all the isomers allowed by the molecular formula.
Clearly, this procedure does not \emph{explain} the molecular shape, since this property is not derived from first principles but is instead introduced in a semiempirical manner by presupposing that the nuclei are distinguishable and largely localized \cite{Woolley1976,Woolley1988a,Sutcliffe2012,Sutcliffe2013}.

Taking for granted the BO approximation, an otherwise first-principles method for defining atoms and their bonding in molecules has been presented by Langhoff and coworkers (see Ref. \citenum{Mills2018} and references therein).
Hence, there remains the challenge of clarifying to what extent and under what conditions the molecular shape 
\emph{emerges} \cite{Primas1998} in the first place from a fully ab-initio theory, which treats nuclei and electrons on an equal footing \cite{Woolley1976,Woolley1978,Claverie1980,Primas1983,Woolley1986,Woolley1988a,Martinez2019}.

Two main viewpoints about this `molecular structure conundrum' \cite{Weininger1984,Woolley1985} have been advanced. First, molecular shape is an intrinsic property, i.e., it can be attributed to the isolated system, that either arises only in nonstationary states, in compliance with position-energy complementarity \cite{Woolley1976,Woolley1978}, or emerges in a stationary state from the correlations between all the constituent particles \cite{Claverie1980,Garcia1981,Matyus2011a,Matyus2011b}, or originates from the decoherence induced on the nuclei by their interaction with the electrons \cite{Matyus2021,Cassam2021}.
Second, molecular shape is an extrinsic property, i.e., it either is manifested only when the system is considered together with its environment, as a broken symmetry phenomenon \cite{Woolley1976,Woolley1978,Woolley1988a}, or arises from the decoherence induced by the environment on the entire system \cite{Zhong2016}, or is actualized by the act of measurement performed by an observer on the system \cite{Franklin2021}.
Since this problem is intimately related to the classical limit of quantum mechanics, where the quantum peculiarities of particle delocalization, indistinguishability, and entanglement disappear, and even to the measurement process, both of which remain  unclear \cite{Schlosshauer2007,Fortin2016}, some authors are still pessimistic about the possibility of a complete bottom-up solution \cite{Martinez2019,Fortin2021,Sutcliffe2020}.

This work pursues a less ambitious goal, namely, a fully ab-initio characterization of the transition from atomlike to moleculelike shape in an isolated system in a stationary state, without invoking any internal decoherence mechanism. We focus on the simplest system for which the property of molecular shape can be meaningfully ascribed, namely three particles interacting through Coulomb forces, two of them identical and the other one with the opposite charge, in the nonrelativistic ground state \cite{Matyus2011a,Matyus2011b,Baskerville2016}.

Since, strictly speaking, the ordinary notion of shape cannot be attributed to a quantal object \cite{Woolley1976,Woolley1978,Claverie1980,Primas1983,Weininger1984,Woolley1985,Woolley1986,Woolley1988a,Woolley1988b,Martinez2019,Sutcliffe2020}, we had better begin by explaining what we mean by shape in this context.
In Sec. \ref{sec:shape} we make a critical assessment of a notion commonly adopted in the literature \cite{Garcia1981,Matyus2011a,Matyus2011b,Baskerville2016} and then expound, in qualitative terms, a view that we believe to be more appropriate in general.
Then, in Subsec. \ref{subsec:system shape} we apply such view to the three-particle system, exploiting the marginal-conditional exact factorization (MCEF) of the eigenfunction introduced earlier \cite{Salas2017}.

It may be considered a disadvantage of such `pre-BO' \cite{Matyus2018} treatment that it has no place for a PES \cite{Sutcliffe2012,Sutcliffe2013}, such construct being so useful as an interpretive tool.
In fact, in Ref. \citenum{Salas2017} we showed that an exact nonadiabatic PES (NAPES) \cite{Wilson1979,Hunter1975,Cassam2006,Gidopoulos2014,Abedi2012} can be rigorously constructed \emph{for the electrons} in the He atom, and that it has interpretive power, in particular for making connections with classical structure ideas.
In Subsec. \ref{subsec:NAPES} we apply this development to the three-particle system, which allows us to treat atoms and molecules on the same basis, complementing and enriching the discussion on the emergence of molecular shape.

In Sec. \ref{sec:calculations} we implement our formalism for sequences of realistic (from H$^-$ to H$_2^+$) and model three-particle systems.
As in Refs. \citenum{Matyus2011a,Matyus2011b,Rodriguez2013,Becerra2013} we examine the evolution of the system shape with the mass ratio of the single and double particles, which allows us to provide an alternative, univocal characterization of the transition from atomlike to moleculelike shape.

We close with Sec. \ref{sec:conclusions}, where we provide concluding remarks and perspectives for future developments in this line of work.

\section{\label{sec:shape} The notion of shape for a quantal system}

Since we can attribute the shape of a classical (macroscopic) object ultimately to the spatial distribution of its constituents in 3-dimensional physical space, it may seem fitting to do the same for
a quantal system
\cite{Garcia1981,Coppens1997}.
Such quantity is encoded in the spatial probability density, obtained by marginalizing, i.e., integrating over, the spin variables in the squared modulus of the eigenfunction in the position representation.
However, if the system has $N$ particles and their coordinates are defined with respect to a
laboratory frame the domain of this density will be a $3N$-dimensional configuration space.
If the overall spatial translation is eliminated by a transformation to a space-fixed frame, as usual, the domain of this density will be a $(3N-3)$-dimensional configuration space \cite{Bunker1998}.

According to Garc\'ia-Sucre and Bunge \cite{Garcia1981}, the particle distribution in physical space can be constructed from the one-particle marginal probability densities defined by marginalizing the spatial probability density over the coordinates of all the particles but one,
\begin{equation}\label{eq:1-density}
\rho_i(\vec{q})=\int'd^{3N-6}q'|\Phi(\vec{Q})|^2, 
\end{equation}
where $\vec{Q}$ is a $(3N-3)$-vector containing all the coordinates, $\vec{q}$ is a 3-vector, and the prime indicates that the integral is to be performed over the configuration space that excludes the coordinates of the $i$-th particle.
Naturally, if the system contains a subset of indistinguishable particles their one-particle marginal densities will be identical.
Maxima of one such density will indicate the most probable positions of the particle in physical space. 
Alternatively, instead of the one-particle marginal densities the extracule densities can be utilized, which have a similar interpretation but are defined in a different way employing a reference point, commonly the center of mass \cite{Matyus2011a,Matyus2011b,Baskerville2016,King2016}.
For a state with zero total angular momentum (S state) each maximum will actually constitute a 2-dimensional spherical shell centered at the center of mass \cite{Matyus2011a,Baskerville2016}, due to the invariance of the Hamiltonian to overall spatial rotation \cite{Bunker1998}.
These shells, considered independently, can provide partial information about the shape of the system.
For example, for three-particle systems M\'atyus and coworkers have shown that a cut of the extracule density for the two identical particles is sufficient to characterize the transition from atomic to molecular shape \cite{Matyus2011a,Matyus2011b}.
(Nevertheless, Lude\~na and coworkers have demonstrated that the shape of the extracule density depends on the reference point selected for its definition, leaving a degree of ambiguity \cite{Rodriguez2013,Becerra2013}.)
The particle distribution is then obtained by \cite{Garcia1981}
\begin{equation}\label{eq:N-density}
\rho(\vec{q})=\sum_{i=1}^N\rho_i(\vec{q}),
\end{equation}
which fulfills $\int d^3q\rho(\vec{q})=N$ if $\Phi$ is normalized to 1.
According to the analysis of Garc\'ia-Sucre and Bunge, the set of maxima of this density can be taken to define the shape of the system.
In particular, for a semi-rigid molecule the set of regions of high nuclear density presumably correspond to the familiar ball-and-stick model.
This view overlooks the rotational invariance of the Hamiltonian and, with that, the fact that the shells associated with the maxima in the one-particle densities of several particles can strongly overlap (for an example see Ref. \citenum{Matyus2011b}).
This problem could be dealt with by a transformation to a molecule-fixed frame, albeit at the cost of considerable complication \cite{Sutcliffe1992,Bunker1998}.
In any case, the one-particle densities can have several maxima, causing all the possible nuclear arrangements to be mixed up in the particle distribution, confounding the identification of the different structural isomers.
Hence, such prescription does not work in general.

A generally unambiguous concept of shape is straightforward if we return to the original spatial probability density in $(3N-3)$-dimensional configuration space \cite{Claverie1980}.
Maxima of this function will correspond to relatively likely configurations.
We will refer to the arrangement of such maxima in the configuration space as the \emph{quantal shape} of the system.
For simplicity, let us now focus on S states. The nuisance of the overall rotation can be avoided by
employing, from the start, the interparticle distances as a set of internal coordinates \cite{Hylleraas1929,Hylleraas1964,Frost1962,Frolov1999}, which define a relative configuration space of dimension $\dim=N(N-1)/2$, in terms of which the Hamiltonian is cast automatically into a form manifestly invariant to overall translation, overall rotation, and coordinate inversion.
(The remaining symmetries of the full symmetry group of the Hamiltonian are the permutations of the space and spin coordinates within all subsets of identical particles \cite{Bunker1998}.)
But still, even for a three-particle system ($\dim=3$), the spatial probability density will be too complicated, making its analysis cumbersome.
Therefore, it will be convenient to marginalize a chosen set of internal coordinates to obtain a reduced probability density in a relative configuration space of lower dimension, more amenable to visualization \cite{Claverie1980}.
If this set is chosen judiciously, the shape of the system can still be characterized univocally.

For example, in the case of a molecule with $N_n$ nuclei,
marginalization of all the electron-electron and nucleus-electron distances from the full probability density will yield a purely nuclear probability density in a relative configuration subspace of $\dim=N_n(N_n-1)/2$.
Now, the quantal \emph{molecular} shape can be defined as the arrangement of the maxima of this function in the relative nuclear configuration subspace.
Since each maximum defines a nuclear framework in correspondence to the classical shape of a possible structural isomer, the quantal molecular shape comprises all the possible classical molecular shapes.
(However, note that the correspondence is not necessarily one-to-one.
For one thing, if enantiomers are possible, there will be one maximum associated with every \emph{pair} of them. Due to the invariance of the internal Hamiltonian to coordinate inversion, the eigenfunction can be chosen to have definite positive or negative parity. In that case, such maximum will correspond to the configuration of a superposition of both enantiomers with positive or negative parity. This issue is related to the so-called Hund's paradox \cite{Woolley1976,Fortin2016}.
Furthermore, if the molecule contains indistinguishable nuclei, to one classical shape there will correspond several configurations arising from the permutations of such nuclei.)
The form and breadth of a maximum will contain the information about the natures and amplitudes of the distorsions (`vibrations') of the entire corresponding nuclear framework.
This notion of quantal molecular shape in $N_n(N_n-1)/2$ relative nuclear configuration subspace conforms with what Claverie and Diner call `quantum structure' \cite{Claverie1980}, and is to be contrasted with the classical structure usually depicted as a ball-and-stick cartoon, sometimes with the balls replaced by thermal ellipsoids to represent the anisotropic vibrations \cite{Coppens1997}.
Nevertheless, there still remains the `conundrum' of why only one of the structural or optical isomers is observed in chemical experiments, which has elicited so much debate in the literature
\cite{Woolley1976,Woolley1978,Claverie1980,Primas1983,Weininger1984,Woolley1985,Woolley1986,Woolley1988a,Woolley1988b,Martinez2019,Sutcliffe2020,Primas1998,Franklin2021,Fortin2016,Fortin2021}.

\section{\label{sec:formalism} The marginal-conditional exact factorization formalism}

\subsection{\label{subsec:system shape} The quantal shape of the three-particle system}

Let us now specialize these ideas to our three-particle system.
We consider the
charges $q_1=q_2=-q_3$ (according to charge conjugation invariance, it is immaterial whether $q_3$ is positive or negative).
Thus, the masses $m_1=m_2<m_3$ model an atomlike system, with the identical particles playing the role of the electrons, whereas the masses $m_1=m_2>m_3$ model a moleculelike system, with the identical particles playing the role of the nuclei.
The only symmetry of the Hamiltonian that needs explicit consideration is the permutation of the spatial and spin coordinates of particles 1 and 2.
We assume that these particles are fermions, so that the eigenfunction must be antisymmetric under this operation.

For an S state the internal coordinates are  $r_{12},r_{13},r_{23}$, where $r_{ij}$ is the distance between particles $i$ and $j$ \cite{Hylleraas1929,Hylleraas1964,Frost1962,Frolov1999}. They determine the shape and size of the triangle defined by the positions of the three particles, and are independent, except that they are constrained by the triangle condition $|r_{13}-r_{23}|\leq r_{12}\leq r_{13}+r_{23}$. The volume element of this configuration space is $dV=r_{12}r_{13}r_{23}dr_{12}dr_{13}dr_{23}$.

The eigenfunction can be written as the product of a spatial part, $\Phi(r_{12},r_{13},r_{23})$, and a spin eigenfunction which, in turn, is the product of a spin function
for particles 1 and 2 and a spin
function for particle 3 \cite{Levine2014}.
Evidently, in this case the marginalization of the spin variables in the probability density amounts to working only with the spatial eigenfunction, since the spin eigenfunction is normalized.
Hence, the quantal shape of this system is encoded in the joint distribution function
\begin{equation}\label{eq:dist}
D(r_{12},r_{13},r_{23})=r_{12}r_{13}r_{23}|\Phi(r_{12},r_{13},r_{23})|^2
\end{equation}
(which includes the Jacobian because it is a radial probability density).
We marginalize the distance between the identical particles, $r_{12}$, to obtain a 2-dimensional marginal distribution (MD) for finding particles 1 and 2 at distances $r_{13}$ and $r_{23}$ from particle 3, respectively, \emph{regardless} of $r_{12}$ (here and henceforth, when we speak of the probability of finding two particles at a certain distance, we imply within an infinitesimal neighborhood around that distance),
\begin{equation}\label{eq:mardist2}
D_m(r_{13},r_{23}):=\int_{|r_{13}-r_{23}|}^{r_{13}+r_{23}}dr_{12}r_{12}D(r_{12},r_{13},r_{23}).
\end{equation}
Note that if we marginalized
two of the distances we would obtain a 1-dimensional MD
that obviously could not provide, by itself, information about the shape of the system.
Now, according to Bayes' product rule, the joint distribution can be factorized exactly as
\begin{equation}\label{eq:Bayes}
D(r_{12},r_{13},r_{23})=D_m(r_{13},r_{23})D_c(r_{12}|r_{13},r_{23}),
\end{equation}
where 
%
%
$D_c$ is the conditional distribution (CD) function for finding particles 1 and 2 at a distance $r_{12}$, \emph{provided} they are found at distances $r_{13}$ and $r_{23}$ from particle 3, respectively.
Taking into account the normalization of $D$,
\begin{eqnarray}\label{eq:normdist}
&{}&\int_0^{\infty}dr_{13}r_{13}\int_0^{\infty}dr_{23}r_{23}\nonumber\\
&\times&\int_{|r_{13}-r_{23}|}^{r_{13}+r_{23}}dr_{12}r_{12}D(r_{12},r_{13},r_{23})=1,
\end{eqnarray}
and imposing that $D_m$ be normalized in the $\{r_{13},r_{23}\}$ configuration subspace,
\begin{equation}\label{eq:normmar}
\int_0^{\infty}dr_{13}r_{13}\int_0^{\infty}dr_{23}r_{23}D_m(r_{13},r_{23})=1,
\end{equation}
imply that $D_c$ is \emph{locally} normalized at every point of this subspace,
\begin{equation}\label{eq:normcon}
\int_{|r_{13}-r_{23}|}^{r_{13}+r_{23}}dr_{12}r_{12}D_c(r_{12}|r_{13},r_{23})=1.
\end{equation}
%
%
%
When evaluated at selected points ($r_{13},r_{23}$) the CD becomes a 1-dimensional function that can also aid in the analysis.
(It is worth mentioning that Berry and coworkers have employed 2-dimensional CD's defined in an ad hoc way, by fixing the distance between one electron and the nucleus in the joint distribution, to illustrate that two-valence-electron atoms acquire a moleculelike shape in some states 
\cite{Berry1995}.)

\subsection{\label{subsec:NAPES} The nonadiabatic potential energy surface}

The structure of our nonrelativistic three-particle system in an S state is governed by the internal Schr\"odinger equation \cite{Frost1962,Hylleraas1929,Hylleraas1964,Frolov1999}
\begin{equation}\label{eq:SE}
\hat{H}\Phi(r_{12},r_{13},r_{23})=E\Phi(r_{12},r_{13},r_{23}).
\end{equation}
In atomic units ($\hbar=m_e=e=1$), the spin-free Hamiltonian reads \cite{Frost1962,Frolov1999}
\begin{widetext}
\begin{eqnarray}\label{eq:Hamiltonian}
\hat{H}=&-&\frac{1}{2\mu_{12}}\left ( \frac{\partial^2 }{\partial  r_{12}^2} + \frac{2}{r_{12}}\frac{\partial }{\partial  r_{12}}\right)
-\frac{1}{2\mu_{13}}\left ( \frac{\partial^2 }{\partial r_{13}^2} + \frac{2}{r_{13}}\frac{\partial }{\partial r_{13}}\right)-\frac{1}{2\mu_{23}}\left ( \frac{\partial^2 }{\partial r_{23}^2} + \frac{2}{r_{23}}\frac{\partial }{\partial r_{23}}\right)\nonumber\\
&+&\frac{q_1q_2}{r_{12}}+\frac{q_1q_3}{r_{13}}+\frac{q_2q_3}{r_{23}}\nonumber\\
&-&\frac{1}{m_1}\frac{r_{12}^{2}+r_{13}^{2}-r_{23}^{2}}{2r_{12}r_{13}}\frac{\partial ^{2}}{\partial r_{12}\partial r_{13}}
-\frac{1}{m_2}\frac{r_{23}^{2}+r_{12}^{2}-r_{13}^{2}}{2r_{23}r_{12}}\frac{\partial ^{2}}{\partial r_{23}\partial  r_{12}}
-\frac{1}{m_3}\frac{r_{13}^{2}+r_{23}^{2}-r_{12}^{2}}{2r_{13}r_{23}}\frac{\partial ^{2}}{\partial r_{13}\partial r_{23}},
\end{eqnarray}
\end{widetext}
where $m_i$ and $q_i$ are the mass and charge of particle $i$,
\begin{equation}
\frac{1}{\mu_{ij}}\equiv\frac{1}{m{_i}}+\frac{1}{m{_j}}
\end{equation}
defines the reduced mass of particles $i$ and $j$, and naturally $r_{ij}=r_{ji}$.
Note that the first and second lines contain the contributions of each particle pair to the kinetic energy and the Coulombic potential energy, respectively, while the third line contains the kinetic couplings between the particle pairs.

We aim at constructing a NAPES for the degrees of freedom included in the MD, namely, $r_{13}$ and $r_{23}$.
The procedure is analogous to the one presented in Ref. \citenum{Salas2017}, making the correspondences $r_1\rightarrow r_{13}$, $r_2\rightarrow r_{23}$, $r_{12}\rightarrow r_{12}$, and using the Hamiltonian (\ref{eq:Hamiltonian}) instead of the fixed-nucleus Hamiltonian. Hence, here we present only the most relevant equations and interpretations, and refer the reader to Ref. \citenum{Salas2017} for further details.
Following Hunter \cite{Hunter1975}, we introduce the MCEF of the spatial eigenfunction \cite{Salas2017}
\begin{equation}\label{eq:MCF}
\Phi(r_{12},r_{13},r_{23})=\psi(r_{13},r_{23})\chi(r_{12}|r_{13},r_{23}),
\end{equation}
by defining marginal and conditional amplitudes
\begin{eqnarray}\label{eq:defpsi}
\psi(r_{13},r_{23})&:=& e^{i\theta(r_{13},r_{23})}\nonumber\\ &\times&\left[\int_{|r_{13}-r_{23}|}^{r_{13}+r_{23}}dr_{12}r_{12}|\Phi(r_{12},r_{13},r_{23})|^2\right]^{1/2}\nonumber\\
&\equiv&e^{i\theta(r_{13},r_{23})}\left\langle\Phi|\Phi\right\rangle^{1/2},
\end{eqnarray}
\begin{equation}\label{eq:defchi}
\chi(r_{12}|r_{13},r_{23}):=\frac{\Phi(r_{12},r_{13},r_{23})}{\psi(r_{13},r_{23})},
\end{equation}
where, from here onwards, angular brackets express integrals over $r_{12}$ with the Jacobian $r_{12}$. The phase $\exp[i\theta(r_{13},r_{23})]$, with $\theta$ real, is arbitrary and can be chosen to set the symmetries of $\psi$ and $\chi$ with respect to the exchange $r_{13}\leftrightarrow r_{23}$ \cite{Salas2017,Gidopoulos2014}.
It can be shown that the marginal amplitude must be a nodeless (and zeroless) function \cite{Salas2017,Hunter1981}, hence the conditional amplitude remains finite everywhere. Now we can write
\begin{eqnarray}
D_m(r_{13},r_{23})&=&r_{13}r_{23}|\psi(r_{13},r_{23})|^2,\label{eq:qmardist}\\
D_c(r_{12}|r_{13},r_{23})&=&r_{12}|\chi(r_{12}|r_{13},r_{23})|^2,\label{eq:qcondist}
\end{eqnarray}
which, together with Eqs. (\ref{eq:normdist})-(\ref{eq:normcon}), imply the normalization conditions
\begin{eqnarray}
\int_0^{\infty}dr_{13}r_{13}\int_0^{\infty}dr_{23}r_{23}\left\langle \Phi|\Phi\right\rangle&=&1,\\ \label{eq:normPhi}
\int_0^{\infty}dr_{13}r_{13}\int_0^{\infty}dr_{23}r_{23}|\psi|^2&=&1,\\ \label{eq:normpsi}\left\langle \chi|\chi\right\rangle&=&1. \label{eq:normchi}
\end{eqnarray}
The local normalization of $\chi$, Eq. (\ref{eq:normchi}), guarantees that the MCEF (\ref{eq:MCF}) is unique, within the arbitrary phase mentioned above \cite{Gidopoulos2014}.

To derive the equations that govern $\psi$ and $\chi$ we apply the variational principle, as follows. First, we set up the constrained functional \cite{Salas2017,Gidopoulos2014}
\begin{eqnarray}\label{eq:functional}
F[\Phi]&\equiv&\int_0^{\infty}dr_{13}r_{13}\int_0^{\infty}dr_{23}r_{23}\langle\Phi|\hat{H}|\Phi\rangle\nonumber\\
&-&\int_0^{\infty}dr_{13}r_{13}\int_0^{\infty}dr_{23}r_{23}\lambda(r_{13},r_{23})\left(\left\langle \chi|\chi\right\rangle-1\right)\nonumber\\
&-&\epsilon\left(\int_0^{\infty}dr_{13}r_{13}\int_0^{\infty}dr_{23}r_{23}|\psi|^2-1\right),
\end{eqnarray}
where the first term is the expectation value of the energy, the second term enforces the local normalization of $\chi$, and the third term enforces the normalization of $\psi$, with $\lambda(r_{13},r_{23})$ and $\epsilon$ being Lagrange multipliers. Then we impose the extremization condition $\delta F=0$, which yields \cite{Salas2017}
\begin{equation}\label{eq:mareq}
\left[\hat{T}_{13,23}+U(r_{13},r_{23})\right]\psi(r_{13},r_{23})=E\psi(r_{13},r_{23}),
\end{equation}
\begin{equation}\label{eq:coneq}
\hat{\Omega}[\psi]\chi(r_{12}|r_{13},r_{23})
=U(r_{13},r_{23})\chi(r_{12}|r_{13},r_{23}),
\end{equation}
where
\begin{widetext}
\begin{eqnarray}
\hat{T}_{13,23}&\equiv&-\frac{1}{2\mu_{13}}\left ( \frac{\partial^2 }{\partial r_{13}^2} + \frac{2}{r_{13}}\frac{\partial }{\partial r_{13}}\right)-\frac{1}{2\mu_{23}}\left ( \frac{\partial^2 }{\partial r_{23}^2} + \frac{2}{r_{23}}\frac{\partial }{\partial r_{23}}\right),\\\label{eq:T1323}
\hat{\Omega}[\psi]&\equiv&\hat{H}
-\frac{1}{\mu_{13}}\frac{1}{\psi}\frac{\partial\psi}{\partial r_{13}}\frac{\partial}{\partial r_{13}}
-\frac{1}{\mu_{23}}\frac{1}{\psi}\frac{\partial\psi}{\partial r_{23}}\frac{\partial}{\partial r_{23}}
\nonumber\\
&-&\frac{1}{m_1}\frac{r_{12}^{2}+r_{13}^{2}-r_{23}^{2}}{2r_{12}r_{13}}\left(\frac{1}{\psi}\frac{\partial\psi}{\partial r_{13}}\frac{\partial}{\partial r_{12}}\right)
-\frac{1}{m_2}\frac{r_{12}^{2}+r_{23}^{2}-r_{13}^{2}}{2r_{12}r_{23}}
\left(\frac{1}{\psi}\frac{\partial\psi}{\partial r_{23}}\frac{\partial}{\partial r_{12}}\right)
\nonumber\\
&-&\frac{1}{m_3}\frac{r_{13}^{2}+r_{23}^{2}-r_{12}^{2}}{2r_{13}r_{23}}\left(\frac{1}{\psi}\frac{\partial\psi}{\partial r_{23}}\frac{\partial}{\partial r_{13}}+ \frac{1}{\psi}\frac{\partial\psi}{\partial r_{13}}\frac{\partial}{\partial r_{23}}+\frac{\partial^2\psi }{\partial r_{13}\partial r_{23}}\right),\label{eq:defOmega}
\end{eqnarray}
\end{widetext}
and the notation $\hat{\Omega}[\psi]$ indicates that this operator depends on $\psi(r_{13},r_{23})$. The eigenvalues $E$ and $U(r_{13},r_{23})$ are related to the Lagrange multipliers and to each other by \cite{Salas2017}
\begin{eqnarray}
E=\epsilon&=&\int_{0}^{\infty}dr_1r_1\int_{0}^{\infty}dr_2r_2\lambda(r_{13},r_{23}),\\ \label{eq:defepsilon}
U(r_{13},r_{23})&\equiv&
\frac{\lambda(r_{13},r_{23})}{|\psi(r_{13},r_{23})|^2}-\frac{\hat{T}_{13,23}\psi(r_1,r_2)}{\psi(r_{13},r_{23})}.\label{eq:defU}
\end{eqnarray}

We have conveniently arranged Eqs. (\ref{eq:mareq}) and (\ref{eq:coneq}) so as to expose their analogy with the BO vibrational and clamped-nuclei electronic equations, respectively: $r_{12}$ and $\{r_{13},r_{23}\}$ are analogous to the electronic and the nuclear coordinates, respectively; $\hat{\Omega}[\psi]$ and $\hat{T}_{13,23}$ are analogous to the clampled-nuclei Hamiltonian and the nuclear kinetic energy operator, respectively; $\chi$ and $\psi$ are analogous to the electronic and the vibrational eigenfunctions, respectively; and $U(r_{13},r_{23})$ is analogous to the BO PES, which is the sought-after NAPES. It must be kept in mind that, due to the uniqueness of the MCEF, there can be only one marginal (`vibrational') eigenfunction with total energy $E$ associated with this NAPES, in contrast with the BO case where the same PES can support a number of vibrational states with different total energies.
It is very significant that our two main interpretive tools, the MD and the NAPES, turn out to be rigorously related by Eq. (\ref{eq:mareq}) together with Eq. (\ref{eq:qmardist}). In fact, we can regard Eq. (\ref{eq:mareq}) as a Schr\"odinger equation with potential $U(r_{13},r_{23})$ and eigenfunction  $\sqrt{D_m(r_{13},r_{23})/r_{13},r_{23}}$.
This NAPES can be interpreted as the effective potential acting between the single particle and each of the two identical particles, in the mean field of the relative motion between the two identical particles. We emphasize that such effective potential is exact, in contrast with the BO approximation, where to account for nonadiabatic effects at least two PES's must be coupled.
In particular, for an atomlike system Eq. (\ref{eq:mareq}) has the form of an \emph{exact} central-field model, with $U(r_{13},r_{23})$ being the effective radial potential \cite{Salas2017}.
The definition of our NAPES is in the same spirit as that of Hunter \cite{Hunter1975} and Gross and coworkers \cite{Gidopoulos2014,Abedi2012}, which use the full molecular Hamiltonian, but differs from the one of Wilson \cite{Wilson1979,Cassam2006}, which uses the electronic Hamiltonian.
Besides their use as interpretive tools for structural problems, Gross and coworkers  have used time-dependent NAPES's extensively for the interpretation of coupled electron-nuclear dynamics under time-dependent fields (see, e.g., Ref. \citenum{Abedi2012}).

\section{\label{sec:calculations} Calculations and Discussion}

We studied the atomlike-moleculelike transition of the ground state along a sequence of systems composed of three singly-charged fermions. Such state ought to be a singlet ($S=0$) with respect to the total spin of the two identical particles, which implies that the spatial eigenfunction (\ref{eq:MCF})
must be symmetric with respect to the exchange $r_{13}\leftrightarrow r_{23}$.
We set $\theta\equiv 0$, so that both $\psi$ and $\chi$ are also symmetric [see Eqs. (\ref{eq:defpsi}) and (\ref{eq:defchi})].

We considered the sequence of realistic, albeit most of them exotic, systems H$^-$ ($e^-e^-p^+$), Mu$^-$ ($e^-e^-\mu^+$), Ps$^-$ ($e^-e^-e^+$), Mu$_2^+$ ($\mu^+\mu^+e^-$), H$_2^+$ ($p^+p^+e^-$). [Note that Ps$^-$ and Ps$_2^+$ ($e^+e^+e^-$) are equivalent for our purposes.] He ($e^-e^-\alpha^{2+}$) is not included in this sequence, since the alpha particle is doubly charged. We discussed its NAPES and distribution functions in detail in Ref. \citenum{Salas2017}. Here, we make reference to this system when it helps illuminate the discussion. Since in the sequence ($e^{\pm},\mu^{\pm},p^{\pm}$) the mass changes too abruptly, in order to visualize the transition in a smoother way we also considered a series of model systems within a relatively narrow range of the parameter $R\equiv m_s/m_d$, where $m_s$ and $m_d$ are the masses of the single and double particles, respectively.

From the practical standpoint, solving the nonlinear system of Eqs. (\ref{eq:mareq}) and (\ref{eq:coneq}) does not seem any easier than solving the original Schr\"odinger equation (\ref{eq:SE}) \cite{Jecko2015,Gossel2019}.
Since we use these equations for the purposes of analysis, not for obtaining solutions of the latter, our strategy consists of extracting approximate marginal and conditional amplitudes from a variationally optimized wavefunction, using Eqs. (\ref{eq:defpsi}) and (\ref{eq:defchi}), and then evaluating the NAPES by
\begin{equation}\label{eq:defU2}
U(r_{13},r_{23})=\langle\chi|\hat{\Omega}[\psi]|\chi\rangle,
\end{equation}
in accordance to Eq. (\ref{eq:coneq}).

We employed the trial function of Flores-Riveros and Rivas-Silva \cite{Flores1999}
\begin{eqnarray}\label{eq:Flores}
\Phi(r_{13},r_{23},r_{12})&=&N(e^{-\alpha r_{13}-
\beta r_{23}}+e^{-\beta r_{13}-\alpha r_{23}})\nonumber\\
&\times& e^{-\gamma(r_{12}-u_{0})^{2}}r_{12}^{l},
\end{eqnarray}
where $\alpha,\beta,\gamma,u_0\in\mathbb{R}^+$ and $l\in\mathbb{Z}^{0+}$ are variational parameters, and $N$ is the normalization factor.
We selected this simplistic trial function because it is amenable to a clear physical interpretation.
Namely, for atomlike systems the first factor has the form of the Hylleraas unrestricted (different orbitals for different electrons) ansatz for the two-electron atom \cite{Hylleraas1929}.
On the other hand, to interpret this factor for moleculelike systems let us take $\alpha$ as the larger of the two exponential parameters.
Then, when $\alpha$ is sufficiently larger than $\beta$, so that the exponentials that contain $\beta$ remain close to 1 in the range where the accompanying exponentials containing $\alpha$ decay to nearly zero, the function resembles the simple linear combination of atomic orbitals -- molecular orbital (LCAO-MO) approximation, $1s_1(r_{13})+1s_2(r_{23})$, of the $1\sigma_g$ MO \cite{Levine2014}.
The second factor is a ground-state harmonic-oscillator-type eigenfunction that describes vibrational motion along $r_{12}$ about the equilibrium distance $u_0$. For atomlike systems it plays the role of a Gaussian geminal correlation factor, while for moleculelike systems it represents the harmonic vibration of the nuclei. The third factor introduces anharmonicity into that motion when $l>0$.
Moreover, the evolution of the variational parameters with $R$ aids in the characterization of the transition, as follows. The overall extension of the system is controlled by $\beta$: the smaller the value of $\beta$ the larger the extension.
On the other hand, the relative localization of the two identical particles within a given system is controlled by $\gamma$: the larger $\gamma$ the higher the localization.
We can claim that the degree of this localization, in comparison with the overall extension, is indicative of the atomlike or moleculelike character of the system, since we know that the interectronic motion in the atom is relatively loose, whereas the internuclear motion in the molecule is relatively rigid.
Therefore, we can employ the ratio $\beta/\gamma$ as a diagnostic of the progress of the transition along the sequence of systems. In particular, we expect that this ratio decrease as the system becomes more moleculelike.

A further advantage of the trial function (\ref{eq:Flores}) is that all of our calculations can be done analytically, which we achieved with the aid of Wolfram Mathematica 12.0 \cite{WolframResearchInc.2020}. Nevertheless, the expressions are extremely formidable, so we will not show them here.
\begin{table*}[!htbp]
\begin{ruledtabular}
\begin{center}
\caption{\label{Tab:parameters_rs} Optimized variational parameters in the trial function (\ref{eq:Flores}) for the realistic systems. $\alpha,\beta,\gamma$ and $u_0$ are given in atomic units.}
\begin{tabular}{p{1.0cm} r c c c c c c}
 &\multicolumn{1}{c}{$R=m_s/m_d$}  & $\alpha$ & $\beta$ & $\gamma$ & $\beta/\gamma$ & $u_0$ & $l$\\
\hline
 He & 7294 & 2.20740 & 1.41173 & 0.03253 & 43.4 & 4.45774 & 0 \\
H$^-$ & 1836 & 1.07324 & 0.47322 & 0.00957 & 49.4 & 11.49397 & 0\\
Mu$^-$ & 206.8 & 1.06758 & 0.46882 & 0.00957 & 49.0 & 11.41274 & 0\\
Ps$^-$ & 1.000 & 0.52109 & 0.15209 & 0.00744 & 20.4 & 8.18378 & 0\\
Mu$_2^+$ & 0.004836 & 1.11320 & 0.21525 & 1.32386 & 0.16 & 1.76860 & 2\\
H$_2^+$ & 0.000540 & 1.12988 & 0.21672 & 4.87807 & 0.044 & 1.93988 & 2\\
\end{tabular}
\end{center}
\end{ruledtabular}
\end{table*}

\begin{table*}[!htbp]
\begin{ruledtabular}
\begin{center}
\caption{\label{Tab:energies_rs} Realistic systems. $E_{exact}$ is the `exact' energy (the sources of these values can be found in Ref. \cite{Flores1999}), $\langle E\rangle$ is the energy expectation value, $U_e$, $U^{\ddag}$ and ($r_{13,e},r_{23,e}$) are the energy of the minima, the energy of the saddle point, and the coordinates of a minimum (the coordinates of the other  mininum  are  obtained  by  interchanging  these  values) in the NAPES, respectively, and ($r_{13,m},r_{23,m}$) are the coordinates of the maximum ($r_{13,m}=r_{23,m}$) or one of the maxima ($r_{13,m}\neq r_{23,m}$; the coordinates of the other  maximum  are  obtained  by  interchanging  these values) in the MD. All values are given in atomic units.}
\begin{tabular}{l c c c c l r }
 & {\bf $E_{exact}$} & {\bf $\langle E\rangle$} & $U_e$ & $U^{\ddag}$ &{\bf $r_{13,e}$,  $r_{23,e}$} &{\bf $r_{13,m}$,  $r_{23,m}$} \\
 \hline
He & -2.903 & -2.902 & -7.601 & -7.286 & 0.26 \newline 0.43 & 0.61 \newline 0.61 \\
H$^-$ & -0.527 & -0.525 & -1.594 & -0.992 & 0.48 \newline 12.09 & 1.54 \newline 1.54 \\
Mu$^-$ & -0.525 & -0.523 & -1.586 & -0.986 & 0.49 \newline 12.03 & 1.55 \newline 1.55 \\
Ps$^-$ & -0.262 & -0.260 & -0.792 & -0.416 & 0.99 \newline 11.72 & 3.26 \newline 3.31 \\	
Mu$_2^{+}$ & -0.585 & -0.583 & -3.829 & -2.065 & 0.35 \newline 2.20 & 0.76 \newline 1.99  \\
H$_2^{+}$ & -0.597 & -0.596 & -7.544 & -4.484 & 0.25 \newline 2.05 & 0.75 \newline 1.71 \\
\end{tabular}
\end{center}
\end{ruledtabular}
\end{table*}

Table \ref{Tab:parameters_rs} displays the optimal values of the variational parameters for the sequence of realistic systems (He is included for comparison purposes).
It is seen that for the atomic systems Mu$^-$ and H$^-$ ($R\gg 1$) $\alpha\simeq 2\beta$, with $\alpha>1$. A common interpretation of this relationship is that the atom has an inner electron suffering a negative screening, which makes it more strongly bound than the outer electron suffering a positive screening \cite{Hylleraas1929,King2016}.
On the other hand, for the molecular systems Mu$_2^+$ and H$_2^+$ ($R\ll 1$)  $\alpha\simeq 5\beta$, in accordance to the LCAO-MO interpretation given above.
(For H$_2^+$, it is pertinent to note that our results for $\alpha$, $\beta$, and $u_0$ are close to the values $\alpha=1.24$ bohr$^{-1}$, $\beta\equiv 0$, and $R_e=2.00$ bohr obtained from the simple LCAO-MO treatment \cite{Levine2014}.)
Finally, for Ps$^-$ ($R=1)$ $\alpha\simeq 3.4\beta$, so this system can be expected to be somewhat intermediate between atomic and molecular.
In addition, for systems with $R>1$ or $R<1$, it is obtained that $l=0$ or $l=2$, respectively, which means that in the atomic systems the interelectronic motion is harmonic, whereas in the molecular systems the internuclear motion is anharmonic.
Furthermore, it is seen that along the sequence $\beta/\gamma$ decreases (recall that He does not belong in this sequence), as predicted above, and that for the atomic systems $\beta/\gamma\gg 1$, whereas for the molecular systems $\beta/\gamma\ll 1$.

Table \ref{Tab:energies_rs} presents the expectation value of the energy obtained from the trial function (\ref{eq:Flores}) and the `exact' nonrelativistic energy for the realistic systems. The level of agreement is sufficient for our largely qualitative analysis.
Employing more accurate trial functions (see, e.g., Ref. \citenum{Baskerville2016}) does not produce any relevant qualitative changes in the features of the NAPES's, MD's and CD's to be discussed below, as demonstrated in Ref. \citenum{Salas2017} (a similar observation was made for the extracule densities in Ref. \citenum{Matyus2011b}), but might instead obscure the interpretation.

\begin{table}[!htbp]
\begin{ruledtabular}
\begin{center}
\caption{\label{Tab:parameters_ms} Optimized variational parameters in the trial function (\ref{eq:Flores}) for the model systems. $\alpha$, $\beta$, $\gamma$ and $u_0$ are given in atomic units. $l=0$ in all cases.}
\begin{tabular}{c c c c r}
$R=m_s$ & $\alpha$ & $\beta$ & $\gamma$ &\multicolumn{1}{c}{$u_0$}\\
\hline
2.000 & 0.69797 & 0.23174 & 0.00850 & 8.20320\\	
1.750 & 0.66546 & 0.21561 & 0.00836 & 8.05929\\
1.500 & 0.62658 & 0.19699 & 0.00810 & 8.02482\\
1.250 & 0.57969 & 0.17637 & 0.00801 & 7.95740\\
1.000 & 0.52109 & 0.15209 & 0.00744 & 8.18378\\
0.750 & 0.44661 & 0.12358 & 0.00685 & 8.59581\\
0.500 & 0.34870 & 0.08771 & 0.00608 & 9.49008\\
0.375 & 0.28541 & 0.07078 & 0.00501 & 11.38755\\
0.250 & 0.21056 & 0.04927 & 0.00390 & 14.45530\\
0.150 & 0.13915 & 0.03016 & 0.00266 & 19.74359\\
\end{tabular}
\end{center}
\end{ruledtabular}
\end{table}
Table \ref{Tab:parameters_ms} shows the optimal values of the variational parameters for the sequence of model systems, characterized by $R=m_s$ since the mass of the double particle was kept at $m_d=1$ atomic unit. It is observed that as $R$ goes from 2.000 to 0.150 (a much smaller range than the one considered in Table \ref{Tab:parameters_rs}), the ratio $\alpha/\beta$ goes smoothly from 3.0 to 4.6, the same increasing trend seen in Table \ref{Tab:parameters_rs}, associated with the transition from atomlike to moleculelike shape.
In addition, for the atomlike systems ($R>1$) the equilibrium `interelectronic' distance ($u_0$) increases slightly with the mass of the `nucleus' ($m_s$). This is due to the fact that as the nucleus becomes less mobile it is a little less effective at screening the electrons from one another.
On the other hand, for the moleculelike systems ($R<1$) the equilibrium `internuclear' distance ($u_0$) decreases rather strongly with the mass of the `electron' ($m_s$). This is due to the fact that as the electron becomes less mobile, it is more effective at binding the nuclei.
Moreover, it can be seen that across the transition $\beta/\gamma$ decreases monotonically (from 27.3 to 11.3), as expected from the previous discussion.

Table \ref{Tab:energies_ms} displays the expectation value of the energy for the model systems. It is observed that as the mass of the single particle decreases $\langle E\rangle$ increases monotonically, i.e., the system becomes less stable.
This can be attributed to the increase in the contribution of the single particle to the kinetic energy [see Eq. (\ref{eq:Hamiltonian})].

\begin{table}[!htbp]
\begin{ruledtabular}
\begin{center}
\caption{\label{Tab:energies_ms} Model systems. $\langle E\rangle$ is the energy expectation value, ($r_{13,m},r_{23,m}$) are the coordinates of the maximum ($r_{13,m}=r_{23,m}$) or one of the maxima ($r_{13,m}\neq r_{23,m}$; the coordinates of the other maximum are obtained by interchanging these values) in the MD, and SD is the second derivative at the maximum or at the middle point between the two maxima of the MD cuts of Fig. \ref{fig:cuts}. All values are given in atomic units.}
\begin{tabular}{c c r r}
$R=m_s$ & {$\langle E\rangle$} &{ $r_{13,m}$,  $r_{23,m}$} &\multicolumn{1}{c}{SD} \\
\hline
2.000 & -0.346 & 2.43 \newline 2.43 & -0.002059 \\
1.750 & -0.330 & 2.55 \newline 2.55 & -0.001446 \\
1.500 & -0.311 & 2.72 \newline 2.72 & -0.000821 \\
1.250 & -0.288 & 2.94 \newline 2.94 & -0.000320 \\
1.000 & -0.260 & 3.26 \newline 3.31 & 0.000022 \\
0.750 & -0.223 & 3.39 \newline 4.40 & 0.000250 \\
0.500 & -0.174 & 3.84 \newline 6.47 & 0.000274 \\
0.375 & -0.144 & 4.46 \newline 8.43 & 0.000170 \\
0.250 & -0.106 & 5.73 \newline 11.92 & 0.000071 \\
0.150 & -0.070 & 8.39 \newline 17.44 & 0.000018 \\
\end{tabular}
\end{center}
\end{ruledtabular}
\end{table}

\begin{figure*}[!htbp]
    \centering
    \includegraphics[scale=0.73]{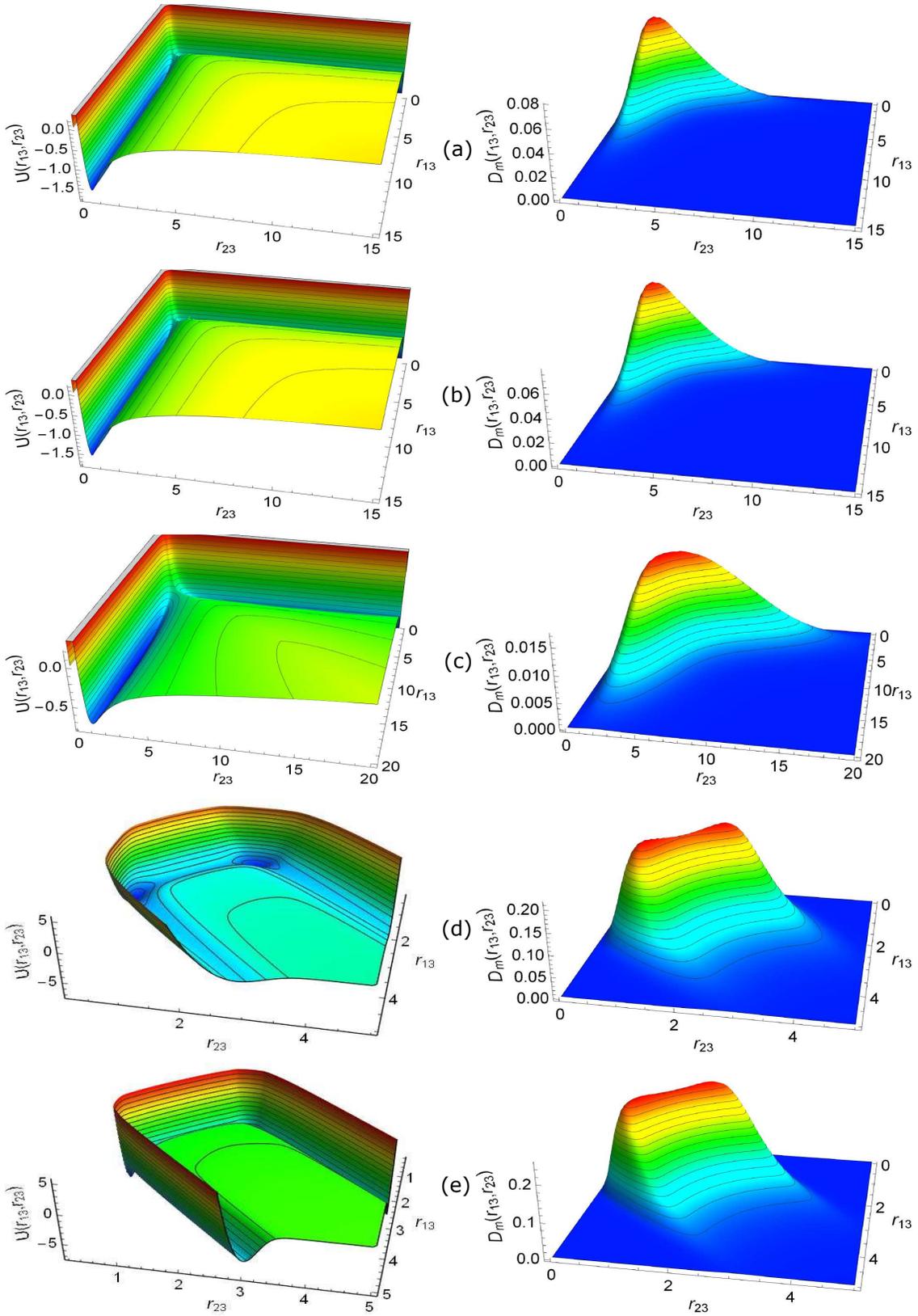}
    \caption{(Color online) NAPES's (left column) and corresponding MD's (right column) for the realistic systems. (a) H$^-$, (b) Mu$^-$, (c) Ps$^-$, (d) Mu$_2^+$, (e) H$_2^+$. All quantities are given in atomic units.}
    \label{fig:RS_U}
\end{figure*}

Figure \ref{fig:RS_U} shows the NAPES's for the sequence of realistic systems and Table \ref{Tab:energies_rs} contains data about the relevant topographical features of these surfaces. It is appreciated that for H$^-$, Mu$^-$ and Ps$^-$ they resemble the one for He discussed in Ref. \citenum{Salas2017}. In particular, they exhibit two perpendicular basins, separated by a ridge along the $r_{13}=r_{23}$ diagonal which goes asymptotically to zero.
A classical interpretation of the topography of such a surface implies that the minima correspond to two broken-symmetry indistinguishable equilibrium configurations (so-called versions \cite{Bunker1998}) with an inner electron and an outer electron \cite{King2016,Hylleraas1929}.
The saddle point that separates the two basins along the mininum potential energy path corresponds to a symmetric configuration with the two electrons on the same orbit.
On the other hand, the NAPES's for Mu$_2^+$ and H$_2^+$ exhibit two nearly parallel basins originating at relatively narrow wells and separated by a plateau centered at the $r_{13}=r_{23}$ diagonal
and that approaches zero asymptotically.
Interestingly, when the internuclear repulsion is removed from $\hat\Omega$ [see Eqs. (\ref{eq:defOmega}) and (\ref{eq:defU2})], these NAPES's change very little, in contrast with the atomic case, where removal of the interelectronic repulsion from $\hat\Omega$ causes the ridge to disappear \cite{Salas2017}.
In a classical interpretation of this topography, the minima correspond to two versions of a broken-symmetry equilibrium configuration, where one of the nuclei orbits the electron more closely than the other. These two minima are separated by a saddle point that corresponds to a symmetric configuration with the two nuclei remaining at the same distance from the electron.
Evidently, for a moleculelike system this NAPES provides explicit information about the relative motion between each nucleus and the electron, with the internuclear motion averaged out, in contrast with the customary BO potential energy \emph{curve}, which provides explicit information about the internuclear motion in the mean field of the electron \cite{Levine2014}.

To envision classical motion over such a NAPES, only trajectories with total energy $\langle E\rangle$ ($\simeq E$) are allowed.
In addition, since the wavefunction (\ref{eq:Flores}) is symmetric with respect to the exchange $r_{13}\leftrightarrow r_{23}$, such motion must be constituted by in-phase breathings of the orbits at all times.
For example, in the atomic case a straight trajectory running along a basin is permitted, as it corresponds to one orbit breathing along $r_{i3}$ with the other one frozen at $r_{j3}=$constant.
In the molecular case, on the other hand, a straight trajectory running along a basin involves an in-phase breathing of both orbits, since the basins run nearly along diagonals, which corresponds to a symmetric vibrational mode.
In both cases, as the total energy lies above the saddle point (see Table \ref{Tab:energies_rs}), the two versions can actually interconvert via trajectories that cross the $r_{13}=r_{23}$ diagonal (a so-called degenerate rearrangement \cite{Bunker1998}).
In the atomic case the asymptotic limit along a basin ($r_{j3}=$constant, $r_{i3}\rightarrow\infty$) would correspond to ionization with the other electron's orbit frozen, while in the molecular case the asymptotic limit along a basin ($r_{13},r_{23}\rightarrow \infty$) would correspond to complete break up of the system, but these limits are in fact unreachable since the state is bound.

\begin{figure}
    \centering
    \includegraphics[scale=0.55]{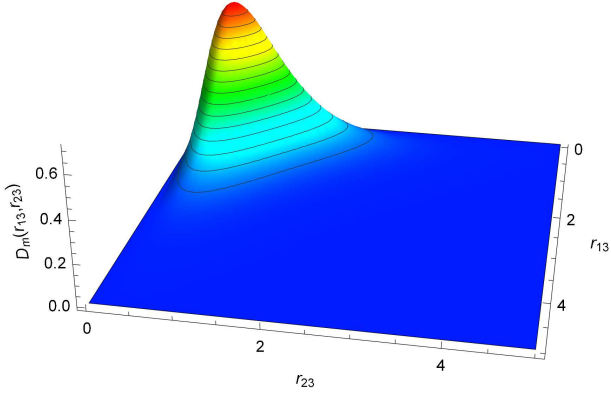}
    \caption{(Color online) MD for the He atom. All quantities are given in atomic units.}
    \label{fig:He_Dmar}
\end{figure}

Figure \ref{fig:RS_U} also displays the corresponding MD's for the realistic systems and Fig. \ref{fig:He_Dmar} shows the MD for the He atom, for comparison purposes. It is helpful to imagine $D_m$ `sitting' on $U$. It is seen that for He, H$^-$, and Mu$^-$ such distribution is unimodal, with its maximum located on the diagonal, i.e. with both electrons on the same orbit, whereas for Mu$_2^+$ and H$_2^+$ such distribution is bimodal, with each maximum located more or less on top of a well (see Table \ref{Tab:energies_rs}).
Hence, the symmetry breaking implied by the NAPES topography is frustrated in the quantal shape of the atomlike systems, that is, there is no inner-outer separation of the electrons, while it is manifested in the quantal shape of the moleculelike systems. (Interestingly, a similar symmetry breaking arises in a fictitious model of He as the \emph{interelectronic repulsion} is enhanced \cite{Salas2017}.)
Evidently, this symmetry breaking in the moleculelike systems is far from sharp, in the sense that configurations where one of the nuclei orbits the electron closer than the other are only slightly preferred over configurations where both nuclei are on the same orbit.
The fact that the probability of finding the electron at the same distance from the two nuclei is less than the probability of finding it closer to one of the nuclei is in accordance with the well-known characteristics of the covalent bond in H$_2^+$ \cite{Levine2014}.
In the case of Ps$^-$, the MD turns out to be slightly bimodal (see Table \ref{Tab:energies_rs}), i.e., this system is more moleculelike than atomlike, despite the topography of its NAPES.
Moreover, the MD for Ps$^-$ is more spread out than the ones of H$^-$ and Mu$^-$, since Ps$^-$ is more weakly bound, and the MD for He is much more compact than the ones for H$^-$ and Mu$^-$, as the nucleus of the former is doubly charged whereas the nuclei of the latter are singly charged.
These results are in general qualitative agreement with the behavior of the extracule densities reported in Ref. \citenum{Matyus2011b}.
We must point out that the symmetry breaking in the classical structure implied by the topography of the NAPES and in the quantal shape of the MD is topological \cite{Matyus2011a}, in contrast with the symmetry breaking associated with a quantum phase transition across a critical point in the eigenspectrum of the system \cite{Shi2001}.

\begin{figure}
    \centering
    \includegraphics[scale=0.92]{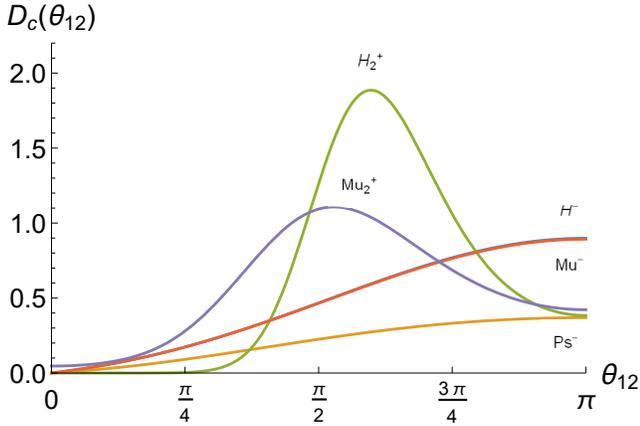}
    \caption{(Color online) CD's for the realistic systems evaluated at the maximum or one of the maxima of the corresponding MD for the atomlike or moleculelike cases, respectively.}
    \label{fig:RS_Dcond_v2}
\end{figure}

Because of the marginalization over $r_{12}$, the NAPES and MD do not provide explicit information about the correlation between the two identical particles.
This kind of information can be retrieved from the CD (\ref{eq:qcondist}).
However, the latter provides radial information, while it is more revealing to examine the angular behavior.
Therefore, we transformed $D_c(r_{12}|r_{13},r_{23})$ into $D_c(\theta_{12}|r_{13},r_{23})$, where $\theta_{12}$ is the angle between the lines joining each of the identical particles with the single particle, by means of the relation $r_{12}=(r_{13}^2+r_{23}^2-2r_{13}r_{23}\cos\theta_{12})^{1/2}$.
Figure \ref{fig:RS_Dcond_v2} shows the CD's for the realistic systems, evaluated at the maximum or one of the maxima ($r_{13,m},r_{23,m}$ in Table \ref{Tab:energies_rs}) of the corresponding MD for the atomic or molecular cases, respectively.
For the atomic systems, H$^-$ and Mu$^-$, the CD's are almost identical and increase monotonically from the collinear nucleus-electron-electron ($\theta_{12}=0$) configuration to the collinear electron-nucleus-electron ($\theta_{12}=\pi$) configuration.
The minimum at $\theta_{12}=0$ reveals the so-called Coulomb hole.
The probability of finding the electrons at zero separation is identically zero because the MD maximum is on the $r_{13}=r_{23}$ diagonal and, hence, the Jacobian present in Eq. (\ref{eq:qcondist}) $r_{12}=0$ when $\theta_{12}=0$.
For the molecular systems, Mu$_2^+$ and H$_2^+$, the CD's have a maximum at $\simeq 90^{\circ}$ and $\simeq 110^{\circ}$, respectively, i.e., at this point in the $\{r_{13},r_{23}\}$  subspace the most probable configuration is bent.
The probability of finding the system in the collinear nucleus-nucleus-electron configuration ($\theta_{12}=0$) now cannot be identically zero because the MD maxima are off the $r_{13}=r_{23}$ diagonal, although for H$_2^+$ it is practically zero.
For Ps$^-$, the CD is monotonic, albeit much flatter than the ones of H$^-$ and Mu$^-$, indicating that this system is more delocalized, in accordance with the larger extension of its MD (see Fig. \ref{fig:RS_U}).
These results are in general qualitative agreement with the angular densities reported in Ref. \citenum{Matyus2011b}, except that there the one of Ps$^-$ appears slightly nonmonotonic, but care must be taken with a direct comparison since those quantities and our CD's are defined differently.
In particular, the CD's shown in Fig. \ref{fig:RS_Dcond_v2} are evaluated at specific points of the $\{r_{13},r_{23}\}$ subspace; their behaviors at other points are different, although not drastically \cite{Salas2017}.

\begin{figure}
    \centering
    \includegraphics[scale=0.59]{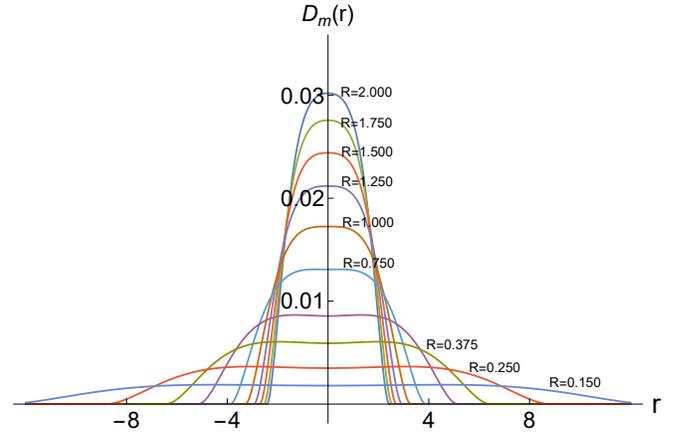}
    \caption{(Color online) Cuts of the MD's through the line perpendicular to the diagonal ($r_{13}=r_{23}$) and passing through the maximum or maxima, for the model systems (see Table \ref{Tab:energies_ms}). $r$ is the distance from the diagonal. All quantities are given in atomic units.}
    \label{fig:cuts}
\end{figure}

\begin{figure}
    \centering
    \includegraphics[scale=0.58]{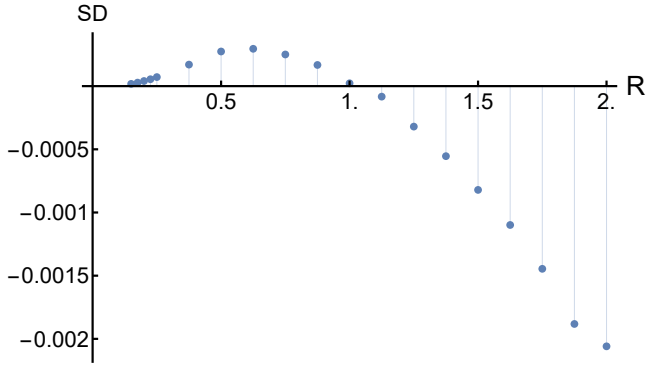}
    \caption{Second derivatives at the middle points of the cuts of Fig. \ref{fig:cuts}.}
    \label{fig:SD}
\end{figure}

In Fig. \ref{fig:RS_U} the change in behavior from atomlike to moleculelike is too sudden, due to the abrupt change of the mass ratio (see Table \ref{Tab:parameters_rs}).
In order to visualize the transition in a smoother fashion
Fig. \ref{fig:cuts} displays cuts of the MD's through the line perpendicular to the diagonal $r_{13}=r_{23}$ and passing through the single maximum ($R>1$) or both maxima ($R<1$), for the series of model systems (see Table \ref{Tab:energies_ms}).
It is seen that as $R=m_s$ decreases such cut becomes more spread out, and that at around $R=1$ it smoothly changes shape from unimodal to bimodal.
To characterize more quantitatively this transition, we evaluated the second derivative (SD) at the middle point of each cut, since when the curvature at this point is negative or positive the distribution is unimodal or bimodal, respectively. The results are presented in Table \ref{Tab:energies_ms} and plotted in Fig. \ref{fig:SD}, where the continuous nature of the transition from atomlike ($R>1$, SD$<0$) to moleculelike ($R<1$, SD$>0$) can be clearly appreciated. The sign change occurs at $R\simeq 1$ (e.g. Ps$^-$), which can be taken to mark the boundary between the two types of behavior.
Below $R\approx 0.6$ the magnitude of SD decreases towards zero, because the cuts become very spread out.

\begin{figure*}[!htbp]
    \centering
    \includegraphics[scale=0.83]{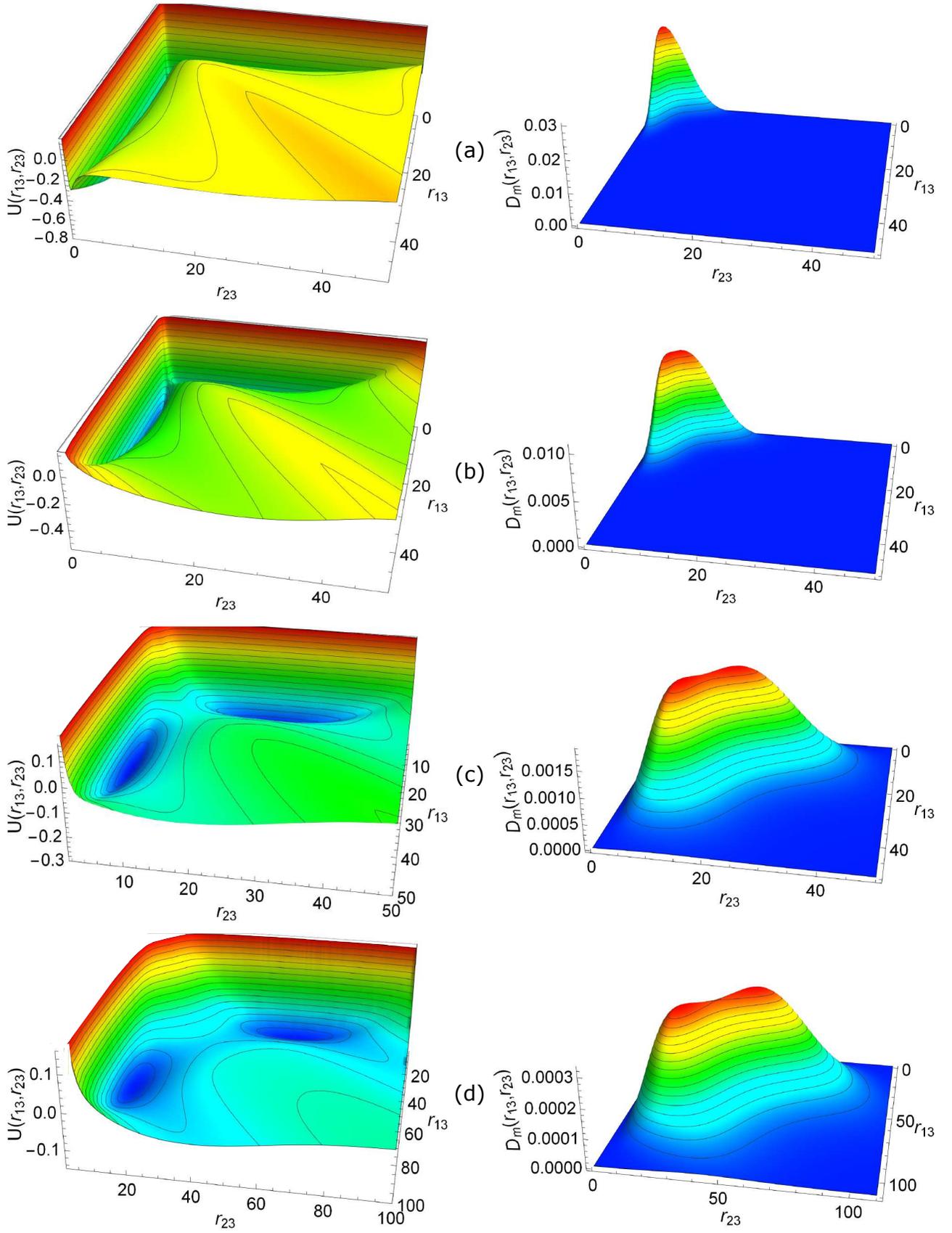}
    \caption{(Color online) NAPES's (left column) and corresponding MD's (right column) for selected model systems. (a) $R=2.00$, (b) $R=0.65$, (c) $R=0.15$, (d) $R=0.05$. All quantities are given in atomic units.}
    \label{fig:RU_surface}
\end{figure*}

In Fig. \ref{fig:RU_surface} we show the NAPES's and the full MD's for selected model systems.
According to Fig. \ref{fig:SD}, $R=2$ is the most atomlike system considered and $R=$0.65, 0.15, 0.05 are moleculelike, with $R=0.65$ possessing the highest SD value.
It is appreciated how the topography of the NAPES gradually deforms, the basins initially being close together and perpendicular, and at the end being more separated and considerably stretched into the diagonal directions (when comparing with Fig. \ref{fig:RS_U} beware of the different scales). 
Correspondingly, the MD gradually becomes more extended, with its shape at the same time changing from unimodal to bimodal, with the humps more clearly differentiated and located on top of the wells in the NAPES.

\section{\label{sec:conclusions} Conclusions and Outlook}


After very briefly reviewing the status of the `molecular structure conundrum' \cite{Woolley1976,Woolley1988a,Woolley1978,Woolley1988b,Claverie1980,Woolley1986,Primas1983,Primas1998,Martinez2019,Weininger1984,Woolley1985,Fortin2016,Franklin2021,Wilson1979,Garcia1981} and stating our view about the concept of shape for an isolated quantal system, we presented an alternative characterization of the continuous emergence of molecular shape in prototypical three-particle Coulomb systems as a response to the variation of the particle masses, solely from the correlations between all the particles treated on an equal footing.

Our analysis was based mainly on the behaviors of the 2-dimensional nonadiabatic potential energy surface and the marginal distribution for the distances between the single particle and each of the two identical particles, constructed rigorously by means of a marginal-conditional exact factorization of a variationally optimized internal wavefunction.
The analysis was further aided by an examination of the evolution of the variational parameters and the conditional distribution for the angle defined by the two identical particles and the single particle.
The classical interpretation of the topography of such surface implies two broken-symmetry indistinguishable equilibrium structures, where both identical particles move around the single particle along separate orbits. Such topological symmetry breaking is frustrated in the atomlike systems but preserved in the moleculelike ones, as revealed by the continuous transition from unimodal to bimodal shape of the quantal marginal distribution.
The conditional distribution, evaluated at selected points of the $\{r_{13},r_{23}\}$ subspace, turns from monotonic to nonmonotonic with a single maximum across the transition.

The unimodal-bimodal transition has already been observed in cuts of the one-particle extracule densities employed in Refs. \citenum{Matyus2011a,Matyus2011b,Baskerville2016}.
Our characterization has two advantages:
First, it is univocal, as it does not rely on the definition of a reference point, in contrast with the one-particle extracule densities, where the bimodality is lost for certain choices of reference points \cite{Rodriguez2013,Becerra2013}.
Second, it permits a unified treatment of atoms, molecules, and any other collection of quantum particles in terms of the potential energy surface concept, which has proven so powerful an interpretive tool in quantum chemistry.


For a three-particle system, three alternative marginal-conditional factorizations are possible that will generate one different 2-dimensional nonadiabatic potential energy surface and two nonadiabatic potential energy \emph{curves}, which, together with their corresponding marginal and conditional distributions, can be employed to discuss related issues of particle correlation and the nature of the Born-Oppenheimer approximation.
In addition, hyperspherical coordinates \cite{Lin1995} constitute an attractive alternative to the interparticle distances, because the hyperradius, $R$, and the angles, $\alpha,\theta_{12}$, are directly related to the size and the shape of the system, respectively.
These investigations are underway in our laboratory.

It has been discovered that certain doubly-excited states of two-electron atoms can be empirically modeled in terms of collective rotational- and bending-like motions of the electrons, analogous to the rovibrational motions of a linear ABA molecule \cite{Kellman1980} (Ref. \citenum{Berry1995} reviews this development).
In addition, it has been predicted that other doubly-excited states of two-electron atoms adopt classical-like configurations with both electrons on the same side of the nucleus (so-called `frozen planet' states) \cite{Richter1992}.
Moreover, it is known that molecules in highly-excited vibrational states can become fluxional. (For a unified review of these three aspects see Ref. \citenum{Kellman1997}).
It is very desirable to develop a unified view of these types of system \cite{Kellman1997}.
We feel that our methodology will be capable of achieving this goal.


\begin{acknowledgments}
Laura D. Salas acknowledges financial support from Colciencias through a doctoral fellowship.
\end{acknowledgments}

\bibliography{Salas_bibliography}

\end{document}